\documentclass[a4paper]{article}
\usepackage{amsmath,theorem,fullpage}
{\theorembodyfont{\rmfamily}
\newtheorem{definition}{Definition}[section]}

{\theorembodyfont{\rmfamily}
\newtheorem{example}{Example}[section]}
\def\Xint#1{\mathchoice
{\XXint\displaystyle\textstyle{#1}}%
{\XXint\textstyle\scriptstyle{#1}}%
{\XXint\scriptstyle\scriptscriptstyle{#1}}%
{\XXint\scriptscriptstyle\scriptscriptstyle{#1}}%
\!\int}
\def\XXint#1#2#3{{\setbox0=\hbox{$#1{#2#3}{\int}$ }
\vcenter{\hbox{$#2#3$ }}\kern-.6\wd0}}
\def\dashint{\Xint-}
\begin{document}
\title{On Vague Computers}
\author{Apostolos Syropoulos\\ 
       Xanthi, Greece\\
    \texttt{asyropoulos@yahoo.com}}
\maketitle
%\authorrunning{Apostolos Syropoulos}
\begin{abstract}
Vagueness is something everyone is familiar with. In fact, most people think
that vagueness is closely related to language and exists only there. However,
vagueness is a property of the physical world. Quantum computers
harness superposition and entanglement to perform their computational tasks.
Both superposition and entanglement are vague processes. Thus quantum
computers, which process exact data without ``exploiting'' vagueness, are
actually vague computers.
\end{abstract} 
%%%%%%%%%%%%%%%%%%%%%%%%%%%%%%%%%%%%%%%%%%%%%%%%%%%%%%%%%%%%%%%%%%%%%%%%%%
\section{Introduction}
%%%%%%%%%%%%%%%%%%%%%%%%%%%%%%%%%%%%%%%%%%%%%%%%%%%%%%%%%%%%%%%%%%%%%%%%%%
Vagueness is something we all are familiar with. A very rough definition of vagueness is this: 
the property of objects or entities that lack definite shape, form, or character.
For many years {\em vagueness} was considered just a linguistic phenomenon.
This simply means that vagueness is part of our everyday expression and not
something real, which turned out not to be true. In the linguistic realm,
a property of some object is vague when it is not clear to which group
or, more generally, category the object belongs to. Thus when Garnet 
is $1{,}68\,\mbox{m}$ tall, it is not obvious if she is tall or not tall. Similarly, 
if Jim is $1{,}90\,\mbox{m}$ tall, he might be classified as tall, in general, but as 
short when his height is compared to the height of the average NBA player. Now,
if John's height is $2{.}00\,\mbox{m}$ and we are sure he is tall, then any other
person whose height is slightly different (e.g., Mike, whose height is $1{.}98\,\mbox{m}$) 
is also considered tall, thus {\em similar} to John with respect to his height. But
what exactly is slightly different? Obviously, this similarity degree is an 
indirect way to define vagueness. In particular, similarity between physical objects 
can be used to show that vagueness exists in the natural world. Although, elementary particles 
of the same kind (i.e., protons) are considered {\em indistinguishable} by most 
physicists, still there are some who argue that elementary particles are in 
fact distinguishable but also similar (e.g., see~\cite{french2008} for an overview). 
Provided this approach is valid, and I will say more on this later on, one can talk 
about vagueness in the physical reality. 

Sometimes people confuse vagueness with {\em ambiguity}. For example, the sentence 
``Garnet ate the cookies on the couch'' is ambiguous because one can understand it
in more than one way. In particular, did Garnet eat the cookies that were on the couch 
or did she bring cookies that she later ate on the couch? Now contrast the previous 
sentence with ``the room was gray,'' which is vague because there are many shades of
gray and it is not clear to which one is the color of the room. People also confuse 
{\em imprecision} with vagueness. For example, the sentence  ``bring me the cup''
is not precise when there are many cups. 
 
A modern computer are fed with exact data, processes them as such and delivers exact 
answers. Of course, this scheme is quite reasonable as we usually have specific 
problems and we want concrete answers to them, but in most, if not all cases, 
we do not care if the internal working involves {\em vague} data and operations 
as long as this does not affect the final result. Of course, it is widely 
assumed that the computational process does not involve any form of vagueness,
yet there are error correction protocols because errors happen. But can we
attribute these errors to vagueness? Furthermore, can we use vagueness constructively 
in the computational process? Or, in different words, is there room for vagueness
in computation? As far as the second question is concerned,, the answer is affirmative 
since there are realistic and not so realistic models of computation that employ vagueness [e.g.,
the fuzzy Turing machine is not a so realistic model, while fuzzy P~systems and fuzzy chemical 
machines are realistic models of computation (see~\cite{syropoulos2014} for details)]. In addition, it seems 
that there is a connection between quantum mechanics and vague computing.

The pillars of modern physics are quantum mechanics and general relativity 
(special relativity explains only the special case where motion is uniform). 
One could say that general relativity is the physics of the macrocosm while 
quantum mechanics is the physics of the microcosm. In different words, one could 
say that quantum mechanics helps us understand the behavior of molecules atoms, and 
elementary particles while general relativity is the theory we use to explain phenomena 
near very massive objects, such as planets, stars, and galaxies (gravity weakens as we go
away from massive objects). It is really weird 
that the two theories do not ``mix''. So far all efforts to quantize gravity have 
failed!\footnote{Speculations about higher dimensions, parallel universes, etc., will 
remain speculations until there is solid proof about their existence.} Quantum mechanics 
started when Max Karl Ernst Ludwig Planck explained the problem of the radiation of a black 
body in 1900. Roughly, he proposed that energy can have only certain discrete values something
that helped him to solve this problem (see~\cite{basdevant2005} among others for a short description 
of the genesis of quantum mechanics).

Quantum computing is making use of the laws of quantum mechanics, and quantum mechanics is
explained by, among others, statistical probabilities (i.e., a combination of statistics
and probability theory). Quantum mechanics was formalized in 1926 while probability theory was 
formalized in 1930~\cite{cook2002}. Until that time, probability theory was considered a prediction tool, 
something that most people still believe. For example, today many people think they can use probabilities 
to make educated bets at a blackjack table and other games of chance. Naturally, they use statistical 
probabilities. Of course, probability theory is not about chance and games. In mathematics, (pure) 
probabilities are {\em ratios of the measure of subsets of a given set}. Here the word ``measure'' 
means ``counting'' in case one deals with finite sets. However, when one has to deal with sets that 
contain an infinite number of elements (e.g., the set of integer numbers), then one must employ a 
suitable measuring process to ``count'' elements. Thus when one knows how to count the elements 
of a set, then one can calculate probabilities. Obviously, this has nothing to do with chance or randomness. 

In what follows I will explore the connection between (mathematical models of) vagueness and quantum
computing. In particular, after a concise introduction to fuzzy set theory, I will introduce possibility
theory. Then I will discuss vagueness at the quantum level and I will explain how possibilities can
replace probabilities in quantum mechanics thus giving rise to real vague computers.

%%%%%%%%%%%%%%%%%%%%%%%%%%%%%%%%%%%%%%%%%%%%%%%%%%%%%%%%%%%%%%%%%%%%%%%%%%
\section{Fuzzy Set Theory: A Mathematical Model of Vagueness}
%%%%%%%%%%%%%%%%%%%%%%%%%%%%%%%%%%%%%%%%%%%%%%%%%%%%%%%%%%%%%%%%%%%%%%%%%%
Fuzzy set theory is a mathematical model of vagueness that was 
introduced by Lotfi Askar Zadeh~\cite{zadeh1965}. Fuzzy sets  are a natural extension of
ordinary sets. Zadeh defined fuzzy sets by generalizing the membership relationship. In 
particular, given a universe $X$, he defined a fuzzy subset of $X$ to be an object that is
characterized by a function $A:X\rightarrow[0,1]$. The value $A(x)$ specifies the degree to 
which an element $x$ belongs to $A$. Thus if $A$ denotes height and $x$ is Garnet, then
$A(x)$ is the degree to which Garnet is tall. A fuzzy set $A$ for which there is
an $x\in X$ such that $A(x)=1$ is called normalized. 

Most newcomers tend to take fuzzy set theory for an alternative formulation 
of probability theory, nevertheless, this is not the case. For instance, there are probability 
theorists that still believe that fuzziness is unnecessary since they argue that probability 
theory can be used to solve all problems that can be tackled by fuzzy set theory. 
Zadeh~\cite{zadeh1995} has argued that the two theories are complementary, that is, they 
are different facets of vagueness. Bart Kosko~\cite{kosko1990} and other researchers, including 
this author~\cite{syropoulos2014}, have argued that fuzzy set theory is more fundamental 
than probability theory. According to Kosko fuzziness ``measures the degree to which an event 
occurs, not whether it occurs. Randomness describes the uncertainty of event occurrence.'' 
However, I do not plan to say anything more on this matter (a very detailed discussion is 
included  in~\cite{syropoulos2014}). Instead, let me now present the basic operations between 
fuzzy subsets.

Assume that $A,B:X\rightarrow[0,1]$ are two fuzzy subsets of $X$. Then, their union and
their intersection are defined as follows:
\begin{align} 
(A\cup B)(x)&=\max\{A(x),B(x)\}\\ 
\intertext{and}
(A\cap B)(x)&=\min\{A(x),B(x)\}.
\end{align}
Also, if $\bar{A}$ is 
the complement of the fuzzy subset $A$, then $\bar{A}(x)=1-A(x)$. More generally, it is quite 
possible to use functions other than $\min$ and $\max$ to define the intersection and the union 
of fuzzy subsets. These functions are known in the literature as {\em t-norms} and {\em t-conorms},
respectively. For more information on t-norms and t-conorms see~\cite{klir95} or any other textbook on
fuzzy set theory.

In the years that followed the publication of Zadeh's paper, various researchers proposed 
and defined various fuzzy structures (e.g., fuzzy algebraic structures, fuzzy topologies, etc.). 
For instance, the concept of fuzzy languages was introduced by E.T. Lee and Zadeh~\cite{lee1969}:
\begin{definition} 
A fuzzy language $\lambda$ over an alphabet $S$ (i.e., an ordinary set of symbols) is 
a fuzzy subset of $S^{\ast}$.
\end{definition}
If $s\in S^{\ast}$, then $\lambda(s)$ is the grade of membership that $s$ is a member of the 
language. 
\begin{example}
Consider the following set that includes all the sequences of zeros followed by ones:
\begin{displaymath}
L = \bigl\{ 0^{i}1^{j}\bigm| i\not=j\;\mbox{and}\, i,j>0\bigr\}.
\end{displaymath}
Then, the following function
\begin{displaymath}
\lambda(0^{i}1^{j})=\left\{ \begin{array}{ll}
                         j/i, & \text{if $i>j$}\\
                         i/j, & \text{otherwise}
                       \end{array}\right.
\end{displaymath}
defines a fuzzy language.
\end{example}

Ordinary set theory is built out of two predicates: membership and equality. This means
that in a fuzzy theory of sets both the membership and the equality should be fuzzy.
Unfortunately, and for unknown reasons, Zadeh {\em fuzzified} only the membership predicate
whereas he left crisp the equality predicate, thus, making the resulting theory somehow
incoherent. It is not difficult to fuzzify the equality predicate and Michael Barr~\cite{barr1986}
has provided a solution to this problem. In addition, he showed how to construct categories
of ``fuzzy'' sets that form a topos. Interestingly, a topos is a non-fuzzy mathematical universe,
thus, he showed how to actually embed ``fuzzy'' sets in such a universe. Although a topos is an
intuitionistic  universe, that is, a universe that is strongly connected to recursion
theory, still it is one that has no respect for vagueness! This implies that it is
necessary to define fuzzy universes, whatever this may mean.

%%%%%%%%%%%%%%%%%%%%%%%%%%%%%%%%%%%%%%%%%%%%%%%%%%%%%%%%%%%%%%%%%%%%%%%%%%
\section{From Probabilities to Possibilities}
%%%%%%%%%%%%%%%%%%%%%%%%%%%%%%%%%%%%%%%%%%%%%%%%%%%%%%%%%%%%%%%%%%%%%%%%%%
Most textbooks on quantum mechanics introduce the reader to the {\em statistical interpretation} of the 
theory in the first pages of the book (e.g., Griffiths's~\cite{griffiths2004} excellent textbook
follows this convention). Of course, the reason is that
the statistical interpretation plays a central r\^{o}le in quantum mechanics. Now, this interpretation 
is based on the pre-Kolmogorov probability theory and uses it for the estimation of the likelihood of 
various events. For example, if $\Psi(x,t)$ is the {\em wave function} of a particle 
that moves on a straight line and $a$ and $b$ are two points of this line, then 
\begin{equation}\label{schrodin}
\int^{b}_{a}|\Psi(x,t)|^2 dx =\left\{\begin{array}{l}
                                      \text{the probability of finding the  par-}\\
                                      \text{ticle between $a$ and $b$, at time $t$.}
                                     \end{array}\right\}
\end{equation}
Obviously, here we are talking about events that we cannot control and so they can be classified
as random. This does not surprise anyone since nonspecialists perceive probabilities as a mathematical
``measure'' of how likely it is to see some event to happen. Statements like the following ones
express exactly this view:
\begin{itemize}
\item it is quite probable that Bayern Munich will win the Champions League this season, or
\item there is a 20\% probability that it will rain tomorrow, or
\item the probability of throwing two dice and obtaining two sixes is 1/36.
\end{itemize}
A rigorous and mathematically sound definition of probabilities have been given by Andrey Nikolaevich 
Kolmogorov in his {\em Analytical Methods of Probability Theory}, which was published in 1931. 
Kolmogorv's formulation appeared almost 6 years after the formalization of quantum mechanics  
(see~\cite{cook2002}). 

Kolmogorov employed measure theory in order to rigorously define probability theory. In particular, 
a probability measure is a function taking sets as arguments that assigns the number $0$ to the empty 
set and a nonnegative number to any other set. Also, it has to be countably additive. Thus given a nonempty
set $X$ and a nonempty class $\mathbf{C}$ of subsets of $X$, and a function $\mu:\mathbf{C}\rightarrow[0,1]$
such that 
\begin{itemize}
\item $\mu(\emptyset)=0$; 
\item $\mu\left(\bigcup_{i=1}^{\infty}E_i\right)=\sum_{i=1}^{\infty}\mu(E_i)$ for any disjoint sequence $\{E_n\}$ of
sets in $\mathbf{C}$ whose union is also in $\mathbf{C}$;
\item  $\mu(X)=1$;
\end{itemize}
then $\mu$ is a probabilistic measure on $\mathbf{C}$. 

As was outlined above, quantum mechanics is using probability theory to explain and predict physical
phenomena. But one could use {\em possibility theory} to give the same explanations and predictions in
a more natural way. In particular, Bart Kosko~\cite{kosko1990}, a prominent fuzzy set theorist, argued in 
favor of the superiority of fuzzy set theory when compared to probability theory by saying that fuzziness 
``measures the degree to which an event occurs, not whether it occurs. Randomness describes the uncertainty 
of event occurrence.'' Thus if the particle lies between $a$ and $b$, we need to know how likely it is 
for the particle to be at $a\le c\le b$ and not whether it is between $a$ and $b$. Possibility theory is 
based on possibility measures, which are are based on fuzzy sets~\cite{zadeh1999}.

A possibility measure $\pi$ is different from a probability measure in that 
\begin{equation}
\pi\left(\bigcup_{i=1}^{\infty}E_i\right)=\sup_{i=1}^{\infty}\pi(E_i).
\end{equation}
In simple words, the difference between the two approaches is that in probability theory one demands that
the sum of probabilities for  given {\em event} should be 1 whereas in possibility theory there should be
at least one plausible event (i.e., one whose possibility is 1). And this is clearly closer to what
actually happens. A particle that lies between $a$ and $b$ is definitely somewhere between them.

Starting from some measure one can define a corresponding integral. For example, when using a
probabilistic measure one may define the  Lebesgue integral. Similarly,
using a possibility measure one can define the {\em Sugeno} integral. 
Assume that $(X, \mathbf{F})$ is measurable space, where $X$ is some set and $\mathbf{F}$ is 
a $\sigma$-algbera,\footnote{$\mathbf{F}$ has
to be a subclass of the power set $2^X$. Also, it must satisfy the following conditions: (a) $X\in\mathbf{F}$; (b) for all 
$E,F\in\mathbf{F}$, $E-F\in\mathbf{F}$;  and (c) for all $E_i\in\mathbf{F}$, $i=1,2,\ldots$, 
$\bigcup_{i=1}^{+\infty}E_{i}\in F$.} $\mu:\mathbf{F}\rightarrow[0,+\infty]$ is continuous {\em monotone} 
measure,\footnote{$\mu$ is monotone if and only if $E,F\in\mathbf{F}$ and $E\subset F$ imply $\mu(E)\le\mu(F)$.}
and $\mathbf{G}$ is the class of all finite nonnegative measurable functions.\footnote{A function 
$f:X\rightarrow(-\infty,+\infty)$ on $X$ is measurable if and only if 
$f^{-1}(B)=\{x\mathrel{|}f(x)\in B\}\in\mathbf{F}$ for any Borel set $B\in\mathcal{B}$.  Now,
assume that $X$ is the real line. Then, the class of all bounded, left closed, and
right open intervals, denoted by $\mathcal{B}$, is the class of Borel sets.}
For any $f\in\mathbf{G}$, $F_{\alpha}=\{x\mathrel{|}f(x)\ge\alpha\}$ and $F_{\alpha^{+}}=\{x\mathrel{|}f(x)>\alpha\}$,
where $\alpha\in[0,+\infty]$. Suppose that $A\in\mathbf{F}$ and $f\in\mathbf{G}$. Then the Sugeno integral
of $f$ on $A$ with respect to $\mu$ is defined by
\begin{equation}
\dashint_{A} fd\mu=\sup_{\alpha\in[0,+\infty]}\Bigl(\alpha\wedge\mu(A\cap F_{\alpha})\Bigr),
\end{equation}
When $A=X$, the Sugeno integral is also denoted by $\dashint fd\mu$. This form of integration
could be used instead of the Lebesgue integral in equation~(\ref{schrodin}) to compute the possibility of 
finding the particle between $a$ and $b$, at time $t$.

%%%%%%%%%%%%%%%%%%%%%%%%%%%%%%%%%%%%%%%%%%%%%%%%%%%%%%%%%%%%%%%%%%%%%%%%%%
\section{Vagueness in the Physical Reality}
%%%%%%%%%%%%%%%%%%%%%%%%%%%%%%%%%%%%%%%%%%%%%%%%%%%%%%%%%%%%%%%%%%%%%%%%%%
If vagueness is not just part of our everyday expression, then there should be vague 
objects. But are there such objects?\footnote{A mathematical response to this question 
has been recently given by Giangiacomo Gerla~\cite{gerla2015}.}
I will not give a ``yes'' or ``no'' answer but
instead I would like to ponder about the length of the UK coastline. The British
Cartographic Society does not give an exact answer on their web page. Instead, they
give this answer: {\em The true answer is: it depends! It depends on the scale at which 
you measure it}. Benoit Mandelbrot~\cite{mandelbrot1967} gave exactly this answer in 1967.
So in a sense it is not exactly known what is inside the UK and what is outside. And of
course it is quite possible that some objects may lie somewhere in the middle. Thus
one could say that the UK is actually a vague object since its boundaries are rigid.
Similarly, clouds are vague objects for exactly the same reasons. On the other hand,
there are objects that appear to be genuine vague objects (e.g., think of heaps of grain or
men with few hair), still most of them are classified as such because the terms that describe 
them are vague. However, there is a third approach to the problem of finding vague 
objects in Nature. In quantum mechanics, the ``standard'' view is that elementary particles
are indistinguishable, nevertheless, not everybody shares this view. More specifically, 
Edward Jonathan Lowe~\cite{lowe1994}, has argued against this view thus showing that 
vagueness exists in the subatomic level:
\begin{quote}\label{vague:real}  
Suppose (to keep matters simple) that in an ionization chamber a
free electron $a$ is captured by a certain atom to form a negative ion which,
a short time later, reverts to a neutral state by releasing an electron $b$. 
As I understand it, according to currently accepted quantum-mechanical    
principles there may simply be no objective fact of the matter as to whether 
or not $a$ is identical with $b$. It should be emphasized that what is 
being proposed here is not merely that we may well have no way of telling whether 
or not $a$ and $b$ are identical,which would imply only an epistemic indeterminacy.
It is well known that the sort of indeterminacy presupposed by orthodox
interpretations of quantum theory is more than merely epistemic---it is
ontic. The key feature of the example is that in such an interaction electron
$a$ and other electrons in the outer shell of the relevant atom enter an 
`entangled' or `superposed' state in which the number of electrons present 
is determinate but the identity of any one of them with $a$ is not, thus 
rendering likewise indeterminate the  identity of $a$ with the released electron
$b$.
\end{quote}
The idea behind this example is that ``identity statements represented by `$a=b$' are `ontically' 
indeterminate in the quantum mechanical context''~\cite{french2003}. In different words, in the 
quantum mechanical context $a$ is equal to $b$ to some degree, which is one of the fundamental 
ideas behind fuzzy set theory. For a thorough discussion of the problem of identity in physics 
see~\cite{french2008}.

%%%%%%%%%%%%%%%%%%%%%%%%%%%%%%%%%%%%%%%%%%%%%%%%%%%%%%%%%%%%%%%%%%%%%%%%%%
\section{Superposition and Entanglement Revisited}
%%%%%%%%%%%%%%%%%%%%%%%%%%%%%%%%%%%%%%%%%%%%%%%%%%%%%%%%%%%%%%%%%%%%%%%%%%
\def\keta#1{\hbox{$\left|#1\right\rangle$}}
\def\ket#1{\hbox{$|\mkern-3mu #1\rangle$}}

The well-known {\em Schr\"{o}dinger's cat paradox} (see~\cite{griffiths2004}) is about a cat that is placed inside 
a box along with a Geiger counter. The box contains a tiny amount of a radioactive substance whose atoms 
may or may not decay within an hour. If there is a decay, it triggers the Geiger counter which, in turn, 
triggers a hammer that breaks a glass that contains a poison capable to kill the cat. The obvious question
is: What would happen to the cat after exactly one hour? At the end of the hour the wave function
of the cat would be
\begin{equation}
\psi=\frac{1}{\sqrt{2}}\psi_{\text{alive}}+\frac{1}{\sqrt{2}}\psi_{\text{dead}}.
\end{equation} 
This implies that the cat is neither dead nor alive! Schr\"{o}dinger regarded this as patent nonsense, however,
I tend to disagree. The reason of course is that there are many things that are not either black or white. After
all, this is exactly the essence of vagueness. Thus, a patient who is in coma is not exactly alive and not
exactly dead. Regardless of our objections, {\em superposition}, that is, the ability of particles to be
in more than one state at the same time, is what makes quantum computing really interesting. 

In ``classical'' computing a bit is either the digit 0 or the digit 1. In quantum computing a {\em qubit} is 
a quantum system (typically a polarized photon, a nuclear spin, etc.) in which the two digits are represented 
by two quantum states: $\keta{0}$ and $\keta{1}$. These states are represented by the following matrices:
\begin{equation}
\keta{0}=\left(\begin{array}{c}1\\ 0\end{array}\right)\;\text{and}\;
\keta{1}=\left(\begin{array}{c}0\\ 1\end{array}\right).
\end{equation}
Also, these two states are ``basic'' states (i.e., they form a basis of a Hilbert space) and any other state of 
the qubit can be written as a superposition $\alpha\keta{0}+\beta\keta{1}$, where $\alpha$ and $\beta$ are complex 
numbers that are called normalization factors and they must obey the normalization condition $|\alpha|^2+|\beta|^2=1$.  
For example, consider a photon that can be polarized in the $x$ direction or in the $y$ direction and assume that
these states are represented by the vectors $\ket{\uparrow}$ and $\ket{\rightarrow}$, respectively, then one can 
use $\ket{\uparrow}$ for $\keta{0}$ and $\ket{\rightarrow}$ for $\keta{1}$. 

The standard interpretation of $\alpha\keta{0}+\beta\keta{1}$ is that a particle is in states $\keta{0}$ or 
$\keta{1}$ with probability that depends on $\alpha$ and $\beta$. Of course, according to a layman's interpretation
of probability theory, these two numbers express the change that a particle is in one of these states. A fuzzy
theoretic interpretation of this state is that the particle is in both  states but with some degree. In fact, 
one can define a fuzzy set as follows:
\begin{align}
\Psi(\keta{0}) &= |\alpha|^2\\
\Psi(\keta{1}) &= |\beta|^2
\end{align}
However, here there is no reason to demand that $|\alpha|^2+|\beta|^2=1$. In fact, there is no reason to impose 
any restriction other than $|\alpha|^2\le1$ and $|\beta|^2\le1$. One may argue that these two restrictions are
not that different, however, the fuzzy theoretic approach assumes that the particle is in fact in a state that is
partly $\keta{0}$ and partly $\keta{1}$. In different words, $\alpha\keta{0}+\beta\keta{1}$ is like a shade of
gray, where, for instance, $\keta{0}$ is like black and $\keta{1}$ is like white. 

Assume $\Psi$ describes the state  of quantum particle in superposition. Then, the superposition collapses upon 
a measurement, but the question is why this happens. Perhaps, the measurement forces a defuzzification of $\Psi$, 
that is, a process by which one gets bivalent data from multivalued data (in this case a vague state is 
transformed into a crisp one). But if defuzzification is possible, then one might expect that fuzzification 
is also possible. Indeed, the Hadamard gate is a mechanism that creates ``vague'' states as follows:
\begin{align}
H \keta{0} &= \frac{1}{\sqrt{2}}\keta{0}+\frac{1}{\sqrt{2}}\keta{1}\\
H \keta{1} &= \frac{1}{\sqrt{2}}\keta{0}-\frac{1}{\sqrt{2}}\keta{1}
\end{align}
Thus superposition corresponds to the fuzzification of a quantum system by means of the $H$ operator,
while measurement is a ``natural'' defuzzification process.

Entanglement is another important quantum mechanical phenomenon. Consider a physical system with two
degrees of freedom, $A$ and $B$. The states of such a system belong to $\mathcal{E}=\mathcal{E}_{A}\otimes
\mathcal{E}_{B}$. Some states can be expressed as 
\begin{equation}
\keta{\Psi}=\keta{\alpha}\otimes\keta{\beta}.
\end{equation}
However, there are states that cannot be {\em factorized} (i.e., they cannot be written as ``products'').
Such states are called {\em entangled states}. For example, the following is such a state:
\begin{equation}
\keta{\Psi} = \frac{1}{\sqrt{2}} \bigl(\keta{\alpha_1} \otimes \keta{\beta_1} + 
                                       \keta{\alpha_2} \otimes \keta{\beta_2}\bigr).
\end{equation}
First of all, there are two ``special'' forms of entanglement, namely entanglement of cost, $E_C$, and 
entanglement of distillation, $E_D$, that vague in a particular case (see~\cite{hwang2003} for details).
More generally, Lowe~\cite{lowe1999} proposed a thought experiment that showed that entanglement is vague.
Assume that there are two determinately distinct electrons. One of them (call it $a$) is determinately
absorbed by an atom and then becomes entangled with a single electron (call it $a^{\ast}$) determinately 
already in the atom. Because these electrons exist in an entangled state inside the atom they are not
determinately distinct but of course we know that there are two of them. At some moment one electron is
emitted and so one electron is still inside the atom and one is outside the atom. Since these two electrons
were in an entangled state,  it is impossible to tell which electron left the atom. In a nutshell, this
is the root of vagueness in entanglement.

Quantum computing is so attractive because it is harnessing both superposition and entanglement
to achieve its exponential computational power. Since both superposition and entanglement are
vague in their nature, this means that quantum computers operate on vague data using vague 
operations. 

%%%%%%%%%%%%%%%%%%%%%%%%%%%%%%%%%%%%%%%%%%%%%%%%%%%%%%%%%%%%%%%%%%%%%%%%%%
\section{Conclusions}
%%%%%%%%%%%%%%%%%%%%%%%%%%%%%%%%%%%%%%%%%%%%%%%%%%%%%%%%%%%%%%%%%%%%%%%%%%
I have briefly explained why vagueness is not only a linguistic phenomenon but also a property of the 
physical world. Also, it is a fact that quantum computers harness quantum mechanical properties of matter 
to perform their computations. These properties of matter have been shown to be vague, thus quantum
computers internally employ vagueness, which makes them automatically vague computers. Of course,
these vague computers process non-vague data in a non-vague way, however, it would be really
interesting to see if processing vague data vaguely would broaden our understanding of computation.
This is certainly an open problem and I think a very interesting one.

%%%%%%%%%%%%%%%%%%%%%%%%%%%%%%%%%%%%%%%%%%%%%%%%%%%%%%%%%%%%%%%%%%%%%%%%%%
\section*{Acknowledgement}
%%%%%%%%%%%%%%%%%%%%%%%%%%%%%%%%%%%%%%%%%%%%%%%%%%%%%%%%%%%%%%%%%%%%%%%%%%
I thank Andromahi Spanou and Christos KK Loverdos for reading the manuscript and helping me to imporve it.

%%%%%%%%%%%%%%%%%%%%%%%%%%%%%%%%%%%%%%%%%%%%%%%%%%%%%%%%%%%%%%%%%%%%%%%%%%

%%%%%%%%%%%%%%%%%%%%%%%%%%%%%%%%%%%%%%%%%%%%%%%%%%%%%%%%%%%%%%%%%%%%%%%%%%

\end{document}